\documentclass[aps,showpacs,superscriptaddress,floatfix]
{revtex4}
\usepackage[dvips]{graphicx}

\newcommand{\ket}[1]{\left | #1 \right \rangle}
\newcommand{\bra}[1]{\left \langle #1 \right |}

\newcommand{\proj}[1]{\ket{#1} \bra{#1}}
\newcommand{\tr}{{\rm \, Tr }\, }
\newcommand{\be}{\begin{equation}}
\newcommand{\ee}{\end{equation}}
\newcommand{\affA}{%
     Communications Research Laboratory,
     Koganei, Tokyo 184-8795, Japan}
\newcommand{\affB}{%
     CREST, Japan Science and Technology Agency}
\newcommand{\affC}{%
     ERATO, Japan Science and Technology Agency}
\newcommand{\affD}{%
      Imai Quantum Computing and Information Project, 
      Bunkyo-ku, Tokyo 113-0033, Japan}
\begin{document}
\title{Quantum learning and universal quantum matching machine}
\author{Masahide Sasaki}
\affiliation{\affA}
\affiliation{\affB}
\author{Alberto Carlini}
\affiliation{\affC}
\affiliation{\affD}
\email{e-mail:psasaki@crl.go.jp; carlini@qci.jst.go.jp}
\begin{abstract}
Suppose that three kinds of quantum systems are given in some
unknown states
$\ket f^{\otimes N}$,
$\ket{g_1}^{\otimes K}$, and
$\ket{g_2}^{\otimes K}$, and
we want to decide which \textit{template} state
$\ket{g_1}$ or $\ket{g_2}$, each representing the feature of the
pattern class ${\cal C}_1$ or ${\cal C}_2$, respectively,
is closest to the input \textit{feature} state $\ket f$.
This is an extension of the pattern matching problem into the quantum domain.
Assuming that these states are known a priori to belong to a certain
parametric family of pure qubit systems, we derive two
kinds of matching strategies.
The first is a semiclassical strategy
which is obtained by the natural extension of conventional matching
strategies and consists of a two-stage procedure:
identification (estimation) of the unknown template states to design the 
classifier
(\textit{learning} process to train the classifier)
and
classification of the input system into the appropriate pattern class
based on the estimated results.
The other is a fully quantum strategy without any intermediate
measurement which we might call as the
{\it universal quantum matching machine}.
We present the Bayes optimal solutions for both strategies in the
case of $K=1$,
showing that there certainly exists a fully quantum matching procedure
which is strictly superior to
the straightforward semiclassical
extension of the conventional matching strategy
based on the learning process.
\end{abstract}
%\pacs{03.67.Hk, 03.65.Ta, 42.50.--p}
% 03.67.Hk Quantum communication
% 03.65.Ta Foundations of quantum mechanics; measurement theory
% 42.50.-p Quantum optics
%
\date{\today}
\maketitle

%%%%%%%%%%%%%%%%%%%%%%%%%%%%%%%%%%%%%%%%%%%%%%%%%%%%%%%%%%%%%
\section{Introduction}\label{introduction}
%%%%%%%%%%%%%%%%%%%%%%%%%%%%%%%%%%%%%%%%%%%%%%%%%%%%%%%%%%%%%

Distinguishing quantum systems is one of the central tasks
in quantum information theory.
We have a useful formalism known as quantum detection and
estimation theory for dealing with this problem
\cite{Helstrom_QDET,Holevo_book,PeresBook}.
Recent progress in quantum communication and computation
provides motivations to generalize this theory and apply it
to various new situations.
Depending on our purposes there may be various scenaria in
the problem of distinguishing quantum systems.
The systems to be distinguished can be sometimes a set of
given quantum states, and sometimes a set of possible
quantum dynamics. These systems are usually generated by a
quantum {\it source} which is expected to have certain
characteristic features. If the source generates a completely
random phenomena, then it is impossible to extract any
meaningful information from it and therefore such a case
will not come into our consideration.
In a broad sense, we may then essentially have three possible circumstances:
\begin{itemize}
\item {(1)}
the source identity, i. e. a set of possible quantum systems and
associated probability distribution, is completely known;
\item{(2)}
the source identity is unknown, but it belongs to
a parametrized family of quantum systems and
probability distributions;
\item{(3)}
the source is known to be stationary and ergodic,
but no other information is available.
\end{itemize}
The case $(1)$ has long been a main subject of quantum
detection and estimation theory.
However, the other two cases are becoming of practical
importance in quantum information technology.
Suppose, for example, we are interested in finding efficient
representations of incoming random sequences of quantum
states.
If the source identity is completely known then
we have well known theorems on the asymptotic average
length of codewords and efficient coding algorithms are being
developed and will be of practical use in the near future.

Consider then the situation in which the source identity is
{\it not} completely known, which is indeed the case
when dealing with a realistic  quantum source.
The obvious way to proceed would be by direct estimation of the source
identity, which is then used in the coding algorithm in place
of the unknown information of the source.
When the source is known to be a member of a parametric
family then the unknown parameters are readily
estimated from the incoming {\it training} data.
With enough data the estimate will be sufficiently
close to the truth and the representation will be nearly
optimal.
On the other hand, if only a limited number of training data
are available, one has to consider an appropriate estimation
strategy which would hopefully be not only asymptotically
optimal when the length of the training set tends to infinity,
but be also optimal for intermediate amounts of training
data.
This kind of problem is known as {\it learning} strategy
in conventional information theory, in particular
in pattern matching theory
\cite{Fu76,Fukunaga90}.
Reasonable criteria which are usually assumed for a good strategy are:
\begin{itemize}
\item{(i)}
no knowledge of the source is required;
\item{(ii)}
the delay due to the learning process is not long;
\item{(iii)}
the strategy should be simple and easy to implement.
\end{itemize}

The purpose of this paper is to develop a formalism for the
quantum learning strategy and to apply it to the problem of
distinguishing quantum systems in cases $(2)$ and $(3)$.
In a recent paper \cite{Sasaki_Carlini_Jozsa2001a},
the authors considered the problem of quantum pattern matching, in which
each pattern class ${\cal C}_i$ is represented by
a known quantum state $\ket{g_i}$ called a template state,
and the task is to find a template which optimally matches
a given unknown quantum state $\ket f$.
Namely we have assumed that
the input states $\ket{f}$ are given as quantum information
(i.e. unknown quantum states) whereas the template states
$\ket{g_i}$'s with known identities are given as classical
information. Our goal was to obtain the best template
as classical information (i.e. knowledge of the identity of the
best $\ket{g_i}$) via a suitable matching strategy which
is represented by a probability operator measure (POM), 
also referred to as a positive operator valued measure (POVM).

In the present paper 
we relax the ingredients of our previous formulation 
in the following way. 
That is,
instead of fully knowing the identities of the template states
we may be given only some finite number ($K$) of copies
of each template
(so our original formulation is equivalent to $K=\infty$).
One matching strategy would then be to apply state estimation to
the sets of $K$ copies and proceed as in our original formulation
with the resulting estimated state identities.
But this is unlikely to be an optimal strategy, since
any intermediate measurement process generally
degrades the classification performance, as shown in Ref.
\cite{Sasaki_Carlini_Jozsa2001a}.
Following the criteria
(i)$\sim$(iii), we should consider
a more fully quantum procedure which, for any input $\ket{f}$,
identifies the best template class
without attempting to obtain any further information
about the identities of the template states themselves.

Unfortunately, however, it seems still difficult to deal with
this problem in general contexts.
Therefore, we mainly consider here some tractable cases in order
to illustrate how the quantum matching strategy should work in general.
In particular, we assume that we {\it a priori} know that
the input feature state $\ket f$ and the template states
$\ket{g_1}$ and $\ket{g_2}$ belong to the following parametrized families
of pure quantum states:
%%%%%%%%%%%%%%%%%%%%
\begin{mathletters}
\begin{eqnarray}
\ket{f}&\equiv&{1\over\sqrt{2}}
              \left(  \ket\uparrow
                      + {\rm e}^{i f} \ket\downarrow
              \right), \\
\ket{g_1}&\equiv&{1\over\sqrt{2}}
              \Big(  \ket\uparrow
                      + {\rm e}^{i g_1} \ket\downarrow
              \Big), \\
\ket{g_2}&\equiv&{1\over\sqrt{2}}
              \Big(  \ket\uparrow
                      + {\rm e}^{i g_2} \ket\downarrow
              \Big),
\label{templates}
\end{eqnarray}
\end{mathletters}
%%%%%%%%%%%%%%%%%%%%
\noindent
where the parameters $f$, $g_1$, and $g_2$ are completely unknown.
In this model, we will compare the semiclassical matching strategy which
is obtained by a natural extension of the conventional matching
strategy, and its fully quantum counterpart
% which works equally well
%for any given states of the above parametrized family of pure quantum states,
which we will identify as the
{\it universal quantum matching machine}.

%%%%%%%%%%%%%%%%%%%%%%%%%%%%%%%%%%%%%%%%%%%%%%%%%%%%%%%%%%%%%
\section{Semiclassical Matching Machine}\label{SCMM}
%%%%%%%%%%%%%%%%%%%%%%%%%%%%%%%%%%%%%%%%%%%%%%%%%%%%%%%%%%%%%

We are now given only some finite number ($K$) of
identical samples of each template $\ket{g_i}$
which represent the features of a class ${\cal C}_i (=1, ..., M)$,
but whose state identities are completely unknown.
The input state $\ket f$ is also given as
an unknown quantum state and
we have $N$ identical copies of $\ket f$.
For simplicity we set $M=2$, i.e., we study the problem of
binary classification.
Thus we start with a system described by the state
\begin{equation}
\ket\Psi\equiv \ket f^{\otimes N}\otimes
\ket{g_1}^{\otimes K}\otimes\ket{g_2}^{\otimes K}.
\label{whole state Psi}
\end{equation}

\begin{figure}
\begin{center}
\includegraphics[width=0.45\textwidth]{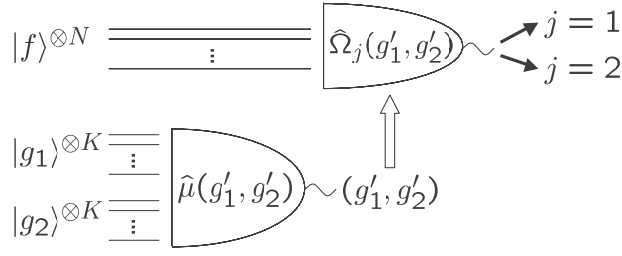}
\end{center}
\caption{\label{fig:LearningScheme}
The semiclassical matching strategy.
The POM $\{\hat\mu(g_1',g_2')\}$ is for estimating the unknown
template states. The output $\{g_1',g_2'\}$ is used to design the
classifier POM.
In other words, using the training data
$\hat g_1^{\otimes K}\otimes\hat g_2^{\otimes K}$,
we fix the classifier to learn the appropriate template parameters.
}
\end{figure}

We first analyze a semiclassical strategy which is a natural
extension of conventional matching strategies.
That is, we 
first apply state estimation to the template states,
design a classifier based on the results,
and then apply this to measure and classify the input feature state.
(see Fig. \ref{fig:LearningScheme}). 
%is the schematic of this semiclassical strategy.
%%%%%%%%%%%%%%%%%%%%
%%%%%%%%%%%%%%%%%%%%
This strategy is represented by two kinds of POMs;
the first is for estimating the identities of the given unknown
template states from the sets of $K$ samples
$\hat g_1^{\otimes K}\otimes\hat g_2^{\otimes K}$
($\hat g_i$ is understood as $\proj{g_i}$).
This POM is indexed by the possible outcomes $\{g_1',g_2'\}$
about the template identities and is denoted by
$\{\hat\mu(g_1',g_2')\}$.
The other is
for classifying the input feature state with $N$ samples
$\hat f^{\otimes N}$.
This POM consists of two elements
$\{\hat\Omega_1(g_1',g_2'), \hat\Omega_2(g_1',g_2')\}$
and should be the optimal matching strategy for the
{\it estimated} templates $\{\hat g_1', \hat g_2'\}$,
which was already given in our previous paper
\cite{Sasaki_Carlini_Jozsa2001a}.
In this way each classifier POM element depends on the estimated
parameters $\{g_1',g_2'\}$.

The problem here is then to find the optimal estimation
strategy $\{\hat\mu(g_1',g_2')\}$.
Such a strategy should maximize the following average
score:
\begin{equation}\label{score by SCMM}
\bar S^{\mathrm{SC}} \equiv\sum_{(g_1',g_2')}\sum_{j=1}^2
\left(  { 1\over{2\pi} }  \right)^3
          \int\int\int_0^{2\pi}dg_1 dg_2 df
\tr\left[ \hat\Omega_j(g_1',g_2')
                        \hat f^{\otimes N} \right]
\tr\left[ \hat\mu(g_1',g_2')
                        \hat g_1^{\otimes K}
                        \otimes\hat g_2^{\otimes K} \right]
\times\vert\langle f\vert g_j\rangle\vert^2.
\end{equation}
The second trace-term in Eq. (\ref{score by SCMM})
is the conditional probability of having
the outcomes $\{g_1',g_2'\}$ for the template states
$\{\hat g_1^{\otimes K}, \hat g_2^{\otimes K}\}$.
The first trace-term is then the conditional probability that
the input state $\hat f$ is classified into the $j$-th class
when an appropriate matching strategy is applied to the $N$
identical input samples $\hat f^{\otimes N}$, and
$\vert\langle f\vert g_j\rangle\vert^2$ is the conditional
score.
%The product of all these terms are summed up over all possibilities

Using the conventional terminology of pattern matching theory,
the POM $\{\hat\mu(g_1',g_2')\}$ corresponds to the {\it learning}
process to train the classifier $\{\hat\Omega_j(g_1',g_2')\}$
with given training samples
$\{\hat g_1^{\otimes K}, \hat g_2^{\otimes K}\}$.
A well known method is the adaptive learning algorithm
in which one first measures each pair of the training samples
$\{\hat g_1, \hat g_2\}$ and then updates the classifier parameters
step by step for $K$ iterations under some appropriate updating
rules.
In contrast, the optimal learning strategy is expected to be a POM
$\{\hat\mu(g_1',g_2')\}$ acting collectively on the state
$\hat g_1^{\otimes K}\otimes\hat g_2^{\otimes K}$,
i.e., fully exploiting the power of quantum entanglement.

The main purpose of this section is to develop
a Bayesian formulation for the optimal learning strategy.
First we introduce the score operators
\begin{equation}\label{score op}
\hat W(g_j)\equiv{1\over{2\pi}}\int_0^{2\pi}df \hat f^{\otimes N}
\vert\langle f\vert g_j\rangle\vert^2,
\end{equation}
just as in our previous paper, and rewrite
Eq. (\ref{score by SCMM}) as
\begin{equation}\label{score by SCMM 2}
\bar S^{\mathrm{SC}} =
       \sum_{(g_1',g_2')}
       \left(  { 1\over{2\pi} }  \right)^2
                  \int\int_0^{2\pi}dg_1 dg_2
      {\tr}\left[ \hat\mu(g_1',g_2')
                             \hat g_1^{\otimes K}
                             \otimes\hat g_2^{\otimes K}
                   \right]
      \sum_{j=1}^2
      {\tr}\left[ \hat\Omega_j(g_1',g_2')
                             \hat W(g_j)
                   \right] .
\end{equation}
We then further introduce {\it a learning score operator}
\begin{equation}\label{def:learning score op}
\hat G(g_1',g_2')\equiv
       \left(  { 1\over{2\pi} }  \right)^2
                  \int\int_0^{2\pi}dg_1 dg_2
                             \hat g_1^{\otimes K}
                             \otimes\hat g_2^{\otimes K}
      \sum_{j=1}^2
      {\tr}\left[ \hat\Omega_j(g_1',g_2')
                             \hat W(g_j)
                   \right] ,
\end{equation}
and rewrite Eq. (\ref{score by SCMM 2}) as
\begin{equation}\label{score by SCMM 3}
\bar S^{\mathrm{SC}} =
       \sum_{(g_1',g_2')}
      {\tr}\left[ \hat\mu(g_1',g_2')
                             \hat G(g_1',g_2')
                   \right] .
\end{equation}
Thus the problem of finding the optimal learning strategy
reduces to the estimation problem of the classifier
parameters $g_1'$ and $g_2'$ through the learning score
operator $\hat G(g_1',g_2')$.

Let us now proceed with the explicit calculation.
We first need to evaluate
$\sum_{j=1}^2
{\tr}\left[
                        \hat\Omega_j(g_1',g_2') \hat W(g_j)
            \right]$.
If the score operator $\hat W(g_j)$ were replaced by
$\hat W(g_j')$, then this quantity would be nothing but the average
score appearing in the quantum
template matching problem discussed in our previous paper
\cite{Sasaki_Carlini_Jozsa2001a}.
In our previous work, the set
$\{\hat\Omega_1, \hat\Omega_2\}$ was designed for
the  {\it a priori} known parameters $g_1$ and $g_2$
of the template states.
On the other hand, the POM
$\{\hat\Omega_1(g_1',g_2'), \hat\Omega_2(g_1',g_2')\}$
here should be designed for the estimated parameters $\{g_1',g_2'\}$,
while the score operators correspond to the unknown template
states $\hat g_1$ or $\hat g_2$.

\begin{figure}
\begin{center}
\includegraphics[width=0.4\textwidth]{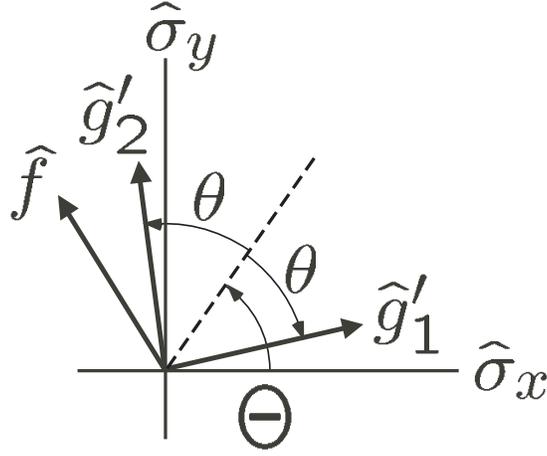}
\end{center}
\caption{\label{fig:StateConfig}
The configuration of the template states in the Bloch sphere
representation.
The input feature state and the template states lie on the
great circle including the x and y axes in the Bloch sphere.
}
\end{figure}

By definition, the POM
$\{\hat\Omega_1(g_1',g_2'), \hat\Omega_2(g_1',g_2')\}$
should maximize the average score for $\hat W(g_j')$
instead of $\hat W(g_j)$, i.e. we should maximiize
\begin{equation}\label{score'}
\bar S'
   = \sum_{j=1}^2 {\tr}
      \left[ \hat W(g_j')\hat\Omega_j(g_1',g_2') \right]
   ={1\over2}+ {\tr}
      \left[ \left(\hat W(g_1') - \hat W(g_2') \right)
                \hat\Omega_1(g_1',g_2') \right],
\end{equation}
where the resolution of the identity
$\hat \Omega_1(g_1',g_2')+\hat \Omega_2(g_1',g_2')=\hat I$
was used in the second equality.
The score operator $\hat \Omega_1(g_1',g_2')$ should be then
taken to maximize
$\tr{\left[ \left(\hat W(g_1')-\hat W(g_2') \right) 
            \hat \Omega(g_1',g_2') \right]}$,
that is,
it should be the projection onto the subspace corresponding to
the positive eigenvalues of the operator $\hat W(g_1')-\hat W(g_2')$.
The score operators are built from  the tensor product of $N$
identical copies of the input system, $\ket f^{\otimes N}$, and
they are most appropriately described on the $N+1$ dimensional
{\it totally symmetric bosonic subspace} of
${\cal H}^{\otimes N}$, ${\cal H}_B$, where $\{\ket m\}$ is the
occupation number basis for the $\downarrow$ component.
The score operators can then be written in the form
\begin{equation}\label{score op2}
\hat W(g_j') = {1\over{2^{N+1}}}
\left[ \sum_{m=0}^N 2 \proj{m}
         +\sum_{m=0}^{N-1}
           \sqrt{ \bigg( \begin{array}{c} N \cr m \end{array}
                      \bigg)
                     \bigg( \begin{array}{c} N \cr m+1 \end{array}
                     \bigg) }
           \left( {\rm e}^{i g_j'}
                       \vert m+1\rangle\langle m\vert
                    +{\rm e}^{-i g_j'}
                       \vert m\rangle\langle m+1\vert
           \right)
\right] .
\end{equation}
Therefore
\begin{equation}\label{delta W}
\hat W(g_1') - \hat W(g_2')
= {{{\sin}\theta}\over{2^{N+1}}}
    \sum_{m=0}^{N-1}
           \sqrt{ \bigg( \begin{array}{c} N \cr m \end{array}
                      \bigg)
                     \bigg( \begin{array}{c} N \cr m+1 \end{array}
                     \bigg) }
           \left( {\rm e}^{i(\Theta+\pi/2)}
                       \vert m+1\rangle\langle m\vert
                    +{\rm e}^{-i(\Theta+\pi/2)}
                       \vert m\rangle\langle m+1\vert
           \right),
\end{equation}
where we have introduced
$\Theta\equiv{{g_1' + g_2'}\over2}$
and
$\theta\equiv{{g_1' - g_2'}\over2}$.
Eq. (\ref{delta W}) can also be rewritten as
\begin{equation}\label{delta W 2}
\Delta\hat W(\Theta+{\pi\over2}, \theta)
\equiv  \hat W(g_1') - \hat W(g_2')
= \hat V(\Theta+{\pi\over2})
    \Delta\hat W(0, \theta)
    \hat V^\dagger(\Theta+{\pi\over2}),
\end{equation}
where
\begin{equation}\label{rotation V}
\hat V(\Theta)
\equiv
\sum_{m=0}^{N} {\rm e}^{im\Theta} \proj{m},
\end{equation}
and
\be
\Delta\hat W(0, \theta)
\equiv
{{{\sin}\theta}\over{2^{N+1}}}
    \sum_{m=0}^{N-1}
           \sqrt{ \bigg( \begin{array}{c} N \cr m \end{array}
                      \bigg)
                     \bigg( \begin{array}{c} N \cr m+1 \end{array}
                     \bigg) }
           \big( \vert m+1\rangle\langle m\vert
                    +\vert m\rangle\langle m+1\vert
           \big).
\ee
Let the spectral decomposition of
$\Delta\hat W(0, \theta)$ be
\be\label{lambda_m}
\Delta\hat W(0, \theta)
=
\sum_{m=0}^{N} \lambda_m\proj{\lambda_m},
\ee
and introduce the POM
\begin{equation}\label{Lambda_1,Lambda_2}
\hat\Lambda_1=\sum_{\lambda_m\ge0} \proj{\lambda_m},
\quad
\hat\Lambda_2=\sum_{\lambda_m<0} \proj{\lambda_m}.
\end{equation}
Note that the $\{\ket{\lambda_m}\}$, and hence the $\{\hat\Lambda_j\}$,
do not depend on $\theta$ while the eigenvalues
$\lambda_m$ are proportional to ${\sin}\theta$.
The optimal strategy for maximizing
$\bar S'$ (Eq. (\ref{score'})) is then expressed by the POM
\begin{equation}\label{Omega_1,Omega_2}
\hat\Omega_j(g_1',g_2')
\equiv \hat\Omega_j(\Theta)
= \hat V(\Theta+{\pi\over2})
    \hat\Lambda_j
    \hat V^\dagger(\Theta+{\pi\over2}).
    \label{omega}
\end{equation}
The parameter $\Theta$ represents the relative position of
the pair of the estimated states $\hat g_1'$ and $\hat g_2'$
from the $\hat\sigma_x$-axis in the Bloch sphere.
This is the only parameter needed to specify the classifier, that is,
the one to be learned from the training samples
$\{\hat g_1^{\otimes K}, \hat g_2^{\otimes K}\}$.
The angle $\theta$ between $\hat g_1'$ and $\hat g_2'$
in the Bloch sphere, on the other hand,  is irrelevant for the
design of the classifier.
The state configuration is depicted in Fig. \ref{fig:StateConfig}.
%%%%%%%%%%%%%%%%%%%%
%%%%%%%%%%%%%%%%%%%%

Using Eq. (\ref{omega}) we then obtain
\begin{eqnarray}\label{score''}
   \sum_{j=1}^2 {\tr}
   \left[ \hat\Omega_j(g_1',g_2') \hat W(g_j) \right]
&=&\sum_{j=1}^2 {\tr}
       \left[ \hat\Omega_j(\Theta) \hat W(g_j) \right]
\nonumber\\
&=&{1\over2} +{\tr}
      \left[ \hat\Lambda_1 
             \Delta\hat W( {{g_1+g_2}\over2}-\Theta,
                                       {{g_1-g_2}\over2} )
      \right]
\nonumber\\
&=&{1\over2}
     +\left[   {\sin}(g_1-\Theta)
                - {\sin}(g_2-\Theta) \right]  R_N,
\end{eqnarray}
where
\begin{equation}\label{R_N}
R_N\equiv{1\over{2^{N+1}}}
    \sum_{m=0}^{N-1}
           \sqrt{ \bigg( \begin{array}{c} N \cr m \end{array}
                      \bigg)
                     \bigg( \begin{array}{c} N \cr m+1 \end{array}
                     \bigg) }
           \langle m\vert\hat\Lambda_1\vert m+1\rangle.
\end{equation}
Therefore the learning score operator in
Eq. (\ref{def:learning score op}) can be also rewritten as
\begin{equation}\label{G}
\hat G(\Theta)
= \left(  { 1\over{2\pi} }  \right)^2
                  \int\int_0^{2\pi}dg_1 dg_2
                             \hat g_1^{\otimes K}
                             \otimes\hat g_2^{\otimes K}
\left\{
{1\over2}
              +\Bigl[   {\sin}(g_1-\Theta)
                         - {\sin}(g_2-\Theta) \Bigr]  R_N
\right\}.
\end{equation}
Then the average score of Eq. (\ref{score by SCMM 3})
finally reads
\begin{equation}\label{score by SCMM 4}
\bar S^{\mathrm{SC}} =
       \sum_{\Theta}
      {\tr}\left[ \hat\mu(\Theta)\hat G(\Theta)
                   \right] .
\end{equation}

After the integration of $g_1$ and $g_2$, the learning score operator
$\hat G(\Theta)$ is represented by
\be
\hat G(\Theta)={1\over2}\hat C\otimes\hat C
+{R_N\over2}
\left[
\hat D(\Theta)\otimes\hat C-\hat C\otimes\hat D(\Theta)
\right],
\ee
where
\begin{eqnarray}
\hat C
&\equiv& {1\over{2^K}}
\sum_{k=0}^K
\bigg( \begin{array}{c} K \cr k \end{array} \bigg)\proj{k},
\\
\hat D(\Theta)
&\equiv&
\frac{i}{2^K}
\sum_{k=0}^{K-1}
           \sqrt{ \biggl( \begin{array}{c} K \cr k \end{array}
                      \bigg)
                     \bigg( \begin{array}{c} K \cr k+1 \end{array}
                     \biggr) }
           \Bigl( {\rm e}^{i \Theta}
                       \vert k+1\rangle\langle k\vert
                    -{\rm e}^{-i \Theta}
                       \vert k\rangle\langle k+1\vert
           \Bigr).
\end{eqnarray}
The basis $\{\ket k\}$ is the symmetric bosonic basis for the system
of $K$
identical copies of the template states $\hat g_1^K$ or $\hat g_2^K$.
Although we have not succeeded yet in deriving the optimal POM
$\{\hat\mu(\Theta)\}$
maximizing the above score $\bar S$ for general $K$, in order to
show how the method works
we present here as an example three different kinds of optimal learning
strategies for the case $K=1$.

The first one is the group covariant continuous POM.
First observe that the spectral decomposition of $\hat G(\Theta)$
is as follows
\be
\hat G(\Theta)=
\frac{(1+2R_N)}{8}\proj{a_+(\Theta)}
+\frac{1}{8}\proj{T}
+\frac{1}{8}\proj{a_0(\Theta)}
+\frac{(1-2R_N)}{8}\proj{a_-(\Theta)},
\ee
where we have introduced
\begin{eqnarray}
\ket{a_+(\Theta)}
&\equiv&{1\over2}
\Bigl(-{\rm e}^{-i(\Theta+{\pi\over2})}\ket{\uparrow\uparrow}
       +\sqrt{2}\ket S
       +{\rm e}^{i(\Theta+{\pi\over2})}\ket{\downarrow\downarrow}
\Bigr),
\\
\ket{a_0(\Theta)}
&\equiv&{1\over2}
\Bigl( {\rm e}^{-i(\Theta+{\pi\over2})}\ket{\uparrow\uparrow}
       +\sqrt{2}\ket S
       -{\rm e}^{i(\Theta+{\pi\over2})}\ket{\downarrow\downarrow}
\Bigr),
\\
\ket{a_-(\Theta)}
&\equiv&{1\over{\sqrt2}}
\Bigl( {\rm e}^{-i(\Theta+{\pi\over2})}\ket{\uparrow\uparrow}
       +{\rm e}^{i(\Theta+{\pi\over2})}\ket{\downarrow\downarrow}
\Bigr),
\\
\ket{T}
&\equiv&{1\over{\sqrt2}}
\Bigl( \ket{\uparrow\downarrow}
       +\ket{\downarrow\uparrow}
\Bigr),
\\
\ket{S}
&\equiv&{1\over{\sqrt2}}
\Bigl( \ket{\uparrow\downarrow}
       -\ket{\downarrow\uparrow}
\Bigr).
\end{eqnarray}
If we symbolically denote
\be
\hat G(\Theta)={1\over8}\tilde G(\Theta)
               \oplus{1\over8}\proj{T},
\ee
the optimal POM can be written as
\be
\hat\mu(\Theta)=\tilde\mu(\Theta)\oplus\proj{T},
\ee
and the average score is given by
\be
\bar S^{\mathrm{SC}}={1\over8}+{1\over8}{\tr}\tilde\Gamma,
\label{score_tilde}
\ee
where
\be
\tilde\Gamma\equiv\frac{1}{2\pi}\int_0^{2\pi}d\Theta
\tilde\mu(\Theta)\tilde G(\Theta).
\ee
So we would like to find the POM $\tilde\mu(\Theta)$ maximizing
${\tr}\tilde\Gamma$.
We can see that the square root measurement based on the maximum
eigenvalue state $\ket{a_+(\Theta)}$ is actually the optimal POM.
In fact, using
\be
\hat a=\frac{1}{2\pi}\int_0^{2\pi}d\Theta
\proj{a_+(\Theta)}
={1\over4}\proj{\uparrow\uparrow}
+{1\over2}\proj{S}
+{1\over4}\proj{\downarrow\downarrow},
\ee
the square root measurement is constructed by
\begin{eqnarray}
\tilde\mu(\Theta)&=&\proj{\tilde\mu(\Theta)},
\label{opt_learning_POM}
\\
\ket{\tilde\mu(\Theta)}
&\equiv&
\hat a^{-{1\over2}}\ket{a_+(\Theta)}
\nonumber
\\
&=&
-{\rm e}^{-i(\Theta+{\pi\over2})}\ket{\uparrow\uparrow}
+\ket S
+{\rm e}^{i(\Theta+{\pi\over2})}\ket{\downarrow\downarrow}.
\label{gcov}
\end{eqnarray}
We then have
\be
\tilde\Gamma=
(1+\sqrt2R_N)\proj{\uparrow\uparrow}
+(1+2\sqrt2R_N)\proj{S}
+(1+\sqrt2R_N)\proj{\downarrow\downarrow}.
\ee
It is almost straightforward to prove that
$\tilde\Gamma-\tilde G(\Theta)\ge0$
(i.e., by seeing the eigenvalues 3, 1, and 0),
and that $[\tilde\Gamma-\tilde G(\Theta)]\tilde\mu(\Theta)=0$.
Thus the POM of Eq. (\ref{opt_learning_POM}) is optimal
\cite{Holevo73_condition,Yuen,Helstrom_QDET,Holevo_book}, and
the maximum average score is
\be
\bar S^{\mathrm{SC}}_{\mathrm{max}}={1\over2}+\frac{R_N}{\sqrt2}.
\ee
The POM in Eq. (\ref{gcov}) is group covariant as specified by 
\begin{eqnarray}
\ket{\tilde\mu(\Theta)}
&=&
\hat V(\Theta)\ket{\tilde\mu(0)}
\\
\hat V(\Theta)
&=&
{\rm e}^{-i\Theta}\proj{\uparrow\uparrow}
+\proj S
+{\rm e}^{i\Theta}\proj{\downarrow\downarrow}.
\end{eqnarray}

The second optimal learning strategy is the discrete version
of the above strategy.
Actually there are many equivalent discrete POMs attaining the
same maximum average score $\bar S_{\mathrm{max}}^{\mathrm{SC}}$.
The strategy requiring the minimum number of outputs is most 
appreciated practically. This can be
directly read from
Eq. (\ref{opt_learning_POM}) as
$\{\tilde\mu(0),
\tilde\mu(\frac{2\pi}{3}),\tilde\mu(\frac{4\pi}{3})\}$.

These two strategies have been derived from quantum estimation theory
in the standard way, that is,
by taking the symmetry of the operator $\hat G(\Theta)$ into account.
On the other hand, we may also derive another solution from
intuitive considerations in the following way.
Since the parameters $g_1$ and $g_2$ specifying the template states 
are completely unknown, the two template states are independent, 
i.e.,  there is no a priori correlation between them, 
and they are just described by the product state 
$\ket{g_1}\otimes\ket{g_2}$.
It might then be sensible to expect that there should exist an optimal
learning strategy based on the separate measurement on each template
state. Yet the relative direction between the two measurements
might be correlated for us to be able to choose the appropriate
classifier $\{\Omega_1(\Theta),\Omega_2(\Theta)\}$.
We may then apply a von Neumann measurement on each template
to know about the state identity.
Let us define the two von Neumann measurements
\begin{eqnarray}
\ket{A_\pm}
&\equiv&
\frac{1}{\sqrt2}\Bigl(\ket\uparrow \pm \ket\downarrow \Bigr),
\\
\ket{B_\pm}
&\equiv&
\frac{1}{\sqrt2}\Bigl(\ket\uparrow \pm i \ket\downarrow \Bigr).
\end{eqnarray}
We can then  show that the four output POM with the corresponding 
guesses for $\Theta$
\be
\begin{array}{lllll}
\ket{\mu(\Theta_0)}&=&\ket{A_+}\otimes\ket{B_+},
\quad\Theta_0&=&-3\pi/4\\
\ket{\mu(\Theta_1)}&=&\ket{A_+}\otimes\ket{B_-},
\quad\Theta_1&=&-\pi/4\\
\ket{\mu(\Theta_2)}&=&\ket{A_-}\otimes\ket{B_+},
\quad\Theta_2&=&3\pi/4\\
\ket{\mu(\Theta_3)}&=&\ket{A_-}\otimes\ket{B_-},
\quad\Theta_3&=&\pi/4.
\end{array}
\label{opt_learning_POM_sep}
\ee
is also an optimal learning strategy.
Actually, it can be seen that
$\sum_{i=0}^3{\tr}
\left[\proj{\mu(\Theta_i)}\hat G(\Theta_i)\right]$
is just the maximum average score $\bar S_{\mathrm{max}}^{\mathrm{SC}}$.
Note that in this case, however, the POM of Eq.
(\ref{opt_learning_POM_sep}) is
no longer group covariant.

%%%%%%%%%%%%%%%%%%%%%%%%%%%%%%%%%%%%%%%%%%%%%%%%%%%%%%%%%%%%%
\section{Universal Quantum Matching Machine}\label{UQMM}
%%%%%%%%%%%%%%%%%%%%%%%%%%%%%%%%%%%%%%%%%%%%%%%%%%%%%%%%%%%%%

The strategies described in the previous section would be a good and
practical matching strategy.
But this is not optimal and there is a more fully quantum
procedure which extracts only the required information,
i. e., the classical information on which class is best matched
with $\ket f$, without attempting to obtain any further
information about the identities of the template states
themselves.
The total system at our hand is now represented by the state
\begin{equation}
\ket\Psi\equiv
\ket f^{\otimes N}\otimes
\ket{g_1}^{\otimes K}\otimes\cdots\otimes\ket{g_M}^{\otimes K}.
\end{equation}
The optimal strategy can then be defined by a
straightforward extension of the Bayesian formulation given in our
previous work \cite{Sasaki_Carlini_Jozsa2001a}, with
the score operators now defined by
\begin{equation}
\hat W_i\equiv
\left(\frac{1}{2\pi}\right)^M
\int dg_1 \cdots \int dg_M \int df \proj\Psi
\times\vert\langle f\vert g_i\rangle\vert^2 P(\hat f),
\label {score_op}
\end{equation}
where $P(\hat f)$ is the {\it a priori} probability distribution
of the input feature parameter (taken here as uniform,
i.e. $P(\hat f)=\frac{1}{2\pi}$).
The new ingredients in the present formulation are just the additional
integrations over the unknown parameters for the template states.
The fully quantum optimal strategy is obtained as
a POM $\{\hat \Pi_i\}$ that maximizes the following average
score
\begin{equation}
\bar S^{\mathrm{QM}}=\sum_i^M \tr{(\hat W_i\hat \Pi_i)}.
\label {average_score_Wj}
\end{equation}
Once parametrized families of input and template states are
specified, the obtained solution is expected to work equally
well for any states belonging to such families by its definition.
In this sense we might call this optimal POM as a
universal quantum matching machine.

\subsection{Example: Two state system with $M=2$ and $K=1$}
\label{Example:two state system}

Although the definition of the universal quantum matching machine
is straightforward, it is in general a difficult task to derive an
explicit expression for the corresponding POM.
Here we present an illustrative example to
demonstrate how the universal quantum matching machine
works and attains a performance which cannot be reached by
any other conventional (semiclassical) matching strategy.

As usual by now, the full input system $\ket f^{\otimes N}$ is
most appropriately described on the $N+1$
dimensional totally symmetric bosonic subspace ${\cal H}_B$ as
\begin{equation}
\ket {f}^{\otimes N}
           =\sum_{k=0}^N
             \sqrt{{1\over2^N}
             \left(\begin{array}{c} N \cr k \end{array}\right)}
             {\rm e}^{ ikf} \ket{k},
\label {ket F}
\end{equation}
where $\{\ket k\}$ is the occupation number basis of the
$\downarrow$-component.
In the case of a binary matching problem ($M=2$)
we have the two score operators
\begin{eqnarray}\label{W1}
\hat W_1={1\over{2^{N+2+2K}}}
&\Bigg[&2
     \sum_{k=0}^N \sum_{m=0}^K \sum_{n=0}^K
     \bigg(\begin{array}{c} N \cr k \end{array}\bigg)
     \bigg(\begin{array}{c} K \cr m \end{array}\bigg)
     \bigg(\begin{array}{c} K \cr n \end{array}\bigg)
     \vert k,m,n\rangle\langle k,m,n\vert
\nonumber \\
&+&\sum_{k=0}^{N-1}
     \sqrt{ \bigg(\begin{array}{c} N \cr k \end{array}\bigg)
               \bigg(\begin{array}{c} N \cr k+1 \end{array}\bigg) }
     \sum_{m=0}^{K-1} \sum_{n=0}^K
     \bigg(\begin{array}{c} K \cr n \end{array}\bigg)
     \sqrt{ \bigg(\begin{array}{c} K \cr m \end{array}\bigg)
               \bigg(\begin{array}{c} K \cr m+1 \end{array}\bigg) }
     \nonumber  \\
&\times&
     \Big(  \vert k+1,m,n\rangle\langle k,m+1,n\vert
             +\vert k,m+1,n\rangle\langle k+1,m,n\vert \Big)
\Bigg],
\end{eqnarray}
\begin{eqnarray}\label{W2}
\hat W_2={1\over{2^{N+2+2K}}}
&\Bigg[&2
     \sum_{k=0}^N \sum_{m=0}^K \sum_{n=0}^K
     \bigg(\begin{array}{c} N \cr k \end{array}\bigg)
     \bigg(\begin{array}{c} K \cr m \end{array}\bigg)
     \bigg(\begin{array}{c} K \cr n \end{array}\bigg)
     \vert k,m,n\rangle\langle k,m,n\vert
\nonumber \\
&+&\sum_{k=0}^{N-1}
     \sqrt{ \bigg(\begin{array}{c} N \cr k \end{array}\bigg)
               \bigg(\begin{array}{c} N \cr k+1 \end{array}\bigg) }
     \sum_{m=0}^{K} \sum_{n=0}^{K-1}
     \bigg(\begin{array}{c} K \cr m \end{array}\bigg)
     \sqrt{ \bigg(\begin{array}{c} K \cr n \end{array}\bigg)
               \bigg(\begin{array}{c} K \cr n+1 \end{array}\bigg) }
\nonumber  \\
&\times&
     \Big(  \vert k+1,m,n\rangle\langle k,m,n+1\vert
             +\vert k,m,n+1\rangle\langle k+1,m,n\vert \Big)
\Bigg],
\end{eqnarray}
where
$\vert k,m,n\rangle
    \equiv
    \vert k\rangle\otimes\vert m\rangle\otimes\vert n\rangle$,
and
$\{\vert k\rangle\}$, $\{\vert m\rangle\}$, and $\{\vert n\rangle\}$
are the occupation number basis of the $\downarrow$-component
for $\ket f^{\otimes N}$, $\ket {g_1}^{\otimes K}$, and
$\ket {g_2}^{\otimes K}$, respectively.
We are to maximize the following quantity
\begin{eqnarray}
\bar S^{\mathrm{QM}}
%&=&\tr{(\hat W_1\hat \Pi_1)}+\tr{(\hat W_2\hat \Pi_2)} \\
             { }&=&
%\tr{(\hat W_2)}
          {1\over 2}+\tr{\left[  (\hat W_1-\hat W_2) \hat \Pi_1
\right]}.
\end{eqnarray}
%where the resolution of the identity
%$\hat \Pi_1+\hat \Pi_2=\hat I$ was used in the second equality.
%Since $\tr{(\hat W_2)}=1/2$,
As already explained in section \ref{SCMM} below Eq. (\ref{score'}),
the problem reduces to finding the
%$\hat \Pi_1$ should be taken to maximize
%$\tr{\left[ (\hat W_1-\hat W_2) \hat \Pi_1 \right]}$, that is,
%it should be the projection onto the
subspace corresponding to
the positive eigenvalues of the operator $\hat W_1-\hat W_2$. 
From Eqs. (\ref{W1}) and (\ref{W2}) we have in the case of $K=1$ 
that,
%\begin{eqnarray}\label{W1-W2}
%\hat W_1-\hat W_2
%%&=&{1\over{2^{N+2+2K}}}
%    \sum_{k=0}^{N-1}
%    \sqrt{ \bigg(\begin{array}{c} N \cr k \end{array}\bigg)
%              \bigg(\begin{array}{c} N \cr k+1 \end{array}\bigg) }
%\nonumber \\
%&\times&\Bigg[
%    \sum_{m=0}^{K-1} \sum_{n=0}^K
%    \sqrt{ \bigg(\begin{array}{c} K \cr m \end{array}\bigg)
%              \bigg(\begin{array}{c} K \cr m+1 \end{array}\bigg) }
%    \bigg(\begin{array}{c} K \cr n \end{array}\bigg)
%    \Big(  \vert k+1,m,n\rangle\langle k,m+1,n\vert
%            +\vert k,m+1,n\rangle\langle k+1,m,n\vert \Big)
%\nonumber \\
%&-&\sum_{m=0}^{K} \sum_{n=0}^{K-1}
%    \bigg(\begin{array}{c} K \cr m \end{array}\bigg)
%    \sqrt{ \bigg(\begin{array}{c} K \cr n \end{array}\bigg)
%              \bigg(\begin{array}{c} K \cr n+1 \end{array}\bigg) }
%    \Big(  \vert k+1,m,n\rangle\langle k,m,n+1\vert
%            +\vert k,m,n+1\rangle\langle k+1,m,n\vert \Big)
%\Bigg].
%\end{eqnarray}
% and derive the optimal POM for
%the universalquantum matching machine.
%Then the above equation reads
\begin{eqnarray}\label{W1-W2 ver2}
\hat W_1-\hat W_2
={{\sqrt2}\over{2^{N+4}}}
     \sum_{k=0}^{N-1}
     \sqrt{ \bigg(\begin{array}{c} N \cr k \end{array}\bigg)
               \bigg(\begin{array}{c} N \cr k+1 \end{array}\bigg) }
\times\Bigg[
&-&\vert k+1,00\rangle\langle k,S\vert
   -   \vert k,S\rangle\langle k+1,00\vert
\nonumber\\
&+&\vert k+1,S\rangle\langle k,11\vert
   +   \vert k,11\rangle\langle k+1,S\vert
\Bigg],
\end{eqnarray}
where the state
$\vert k+1,00\rangle$ is understood as
$\vert k+1\rangle\otimes\vert0\rangle\otimes\vert0\rangle$.
(Note that in the $K=1$ case we have that 
   $\vert S\rangle
    \equiv (\vert01\rangle-\vert10\rangle)/\sqrt2
    =(\vert\uparrow\downarrow\rangle-\vert\downarrow\uparrow\rangle) 
    /\sqrt2$, 
 i.e., 
 the bosonic space for the template states reduces to the
 original one-qubit space.) 
%\footnote{Note that in the $K=1$ case we have that $\vert S\rangle
%   \equiv (\vert01\rangle-\vert10\rangle)/\sqrt2=
%   (\vert\uparrow\downarrow\rangle-\vert\downarrow\uparrow\rangle)/\sqr%t2=
%   $, i.e. the bosonic space for the template states reduces to the
%   original one-qubit space.}
The operator $\hat W_1-\hat W_2$ can be finally arranged into a
  direct sum as
\begin{equation}\label{W1-W2 ver3}
\hat W_1-\hat W_2=\bigoplus_{k=0}^N \Delta\hat W_k,
\end{equation}
where
\begin{eqnarray}\label{delta W_k}
\Delta\hat W_k
     \equiv{{\sqrt2}\over{2^{N+4}}}
     \sum_{k=1}^{N-1}
     \sqrt{ \bigg(\begin{array}{c} N \cr k \end{array}\bigg) }
     \Bigg[
           &-&\sqrt{ \bigg(\begin{array}{c}
                                          N \cr k+1
                                     \end{array}\bigg) }
                 \Big( \vert k+1,00\rangle\langle k,S\vert
                        +\vert k,S\rangle\langle k+1,00\vert \Big)
\nonumber\\
           &+&\sqrt{ \bigg(\begin{array}{c}
                                          N \cr k-1
                                     \end{array}\bigg) }
                 \Big( \vert k,S\rangle\langle k-1,11\vert
                        +\vert k-1,11\rangle\langle k,S\vert \Big)
\Bigg],  \quad (1\le k \le N-1),
\end{eqnarray}
\begin{equation}\label{delta W_0}
\Delta\hat W_0
     \equiv{{\sqrt{2N}}\over{2^{N+4}}}
     \Big(-\vert 1,00\rangle\langle 0,S\vert
             -\vert 0,S\rangle\langle 1,00\vert \Big),
\end{equation}
and
\begin{equation}\label{delta W_N}
\Delta\hat W_N
     \equiv{{\sqrt{2N}}\over{2^{N+4}}}
     \Big( \vert N,S\rangle\langle N-1,11\vert
             +\vert N-1,11\rangle\langle N,S\vert \Big).
\end{equation}
Subsequently, the $\Delta\hat W_k$'s are diagonalized  as
\begin{eqnarray}\label{delta W_k diag}
\Delta\hat W_k
     \equiv{{\sqrt2}\over{2^{N+4}}}
     \sqrt{ \bigg(\begin{array}{c} N \cr k \end{array}\bigg) }
     \sqrt{ \bigg(\begin{array}{c} N \cr k+1 \end{array}\bigg)
              +\bigg(\begin{array}{c} N \cr k-1 \end{array}\bigg) }
     \Big( \vert k_+\rangle\langle k_+\vert
            -\vert k_-\rangle\langle k_-\vert \Big),
\quad (1\le k \le N-1),
\end{eqnarray}
\begin{equation}\label{delta W_0 diag}
\Delta\hat W_0
     \equiv{{\sqrt{2N}}\over{2^{N+4}}}
     \Big( \vert 1_+\rangle\langle 1_+\vert
            -\vert 1_-\rangle\langle 1_-\vert \Big),
\end{equation}
and
\begin{equation}\label{delta W_N diag}
\Delta\hat W_N
     \equiv{{\sqrt{2N}}\over{2^{N+4}}}
     \Big( \vert N+1_+\rangle\langle N+1_+\vert
            -\vert N+1_-\rangle\langle N+1_-\vert \Big),
\end{equation}
where
\begin{equation}\label{ket k}
\ket{k_\pm}\equiv
{
{
  \mp\sqrt{ \bigg(\begin{array}{c} N \cr k+1 \end{array}\bigg) }
  \ket{k+1,00}
  +\sqrt{ \bigg(\begin{array}{c} N \cr k+1 \end{array}\bigg)
             +\bigg(\begin{array}{c} N \cr k-1 \end{array}\bigg) }
  \ket{k,S}
  \pm\sqrt{ \bigg(\begin{array}{c} N \cr k-1 \end{array}\bigg) }
  \ket{k-1,11}
}
\over
{
    {\sqrt2}
     \sqrt{ \bigg(\begin{array}{c} N \cr k+1 \end{array}\bigg)
              +\bigg(\begin{array}{c} N \cr k-1 \end{array}\bigg) }
}
},
\end{equation}
for $1\le k \le N-1$,
\begin{equation}\label{ket 0}
\ket{1_\pm}
    \equiv
    {1\over\sqrt2}
   \left(\mp\ket{1,00}+\ket{0,S}\right),
\end{equation}
and
\begin{equation}\label{ket N}
\ket{N+1_\pm}
    \equiv
    {1\over\sqrt2}
   \left(\pm\ket{N,S}+\ket{N-1,11}\right),
\end{equation}
respectively.
Therefore the optimal matching strategy is described by the POM
\begin{equation}\label{optimal POM}
\hat\Pi_1=\sum_{k=1}^{N+1} \proj{k_+}, \quad
\hat\Pi_2=\sum_{k=1}^{N+1} \proj{k_-},
\end{equation}
and the optimal attainable average score is given by
\begin{equation}\label{max score}
\bar S^{\mathrm{QM}} = {1\over2}
+
\frac{\sqrt2}{2^{N+4}}
\left(
     2{\sqrt N}+\sum_{k=1}^{N-1}
     \sqrt{ \bigg(\begin{array}{c} N \cr k \end{array}\bigg) }
     \sqrt{ \bigg(\begin{array}{c} N \cr k+1 \end{array}\bigg)
              +\bigg(\begin{array}{c} N \cr k-1 \end{array}\bigg) }
\right).
\end{equation}

\begin{figure}
\begin{center}
\includegraphics[width=0.6\textwidth]{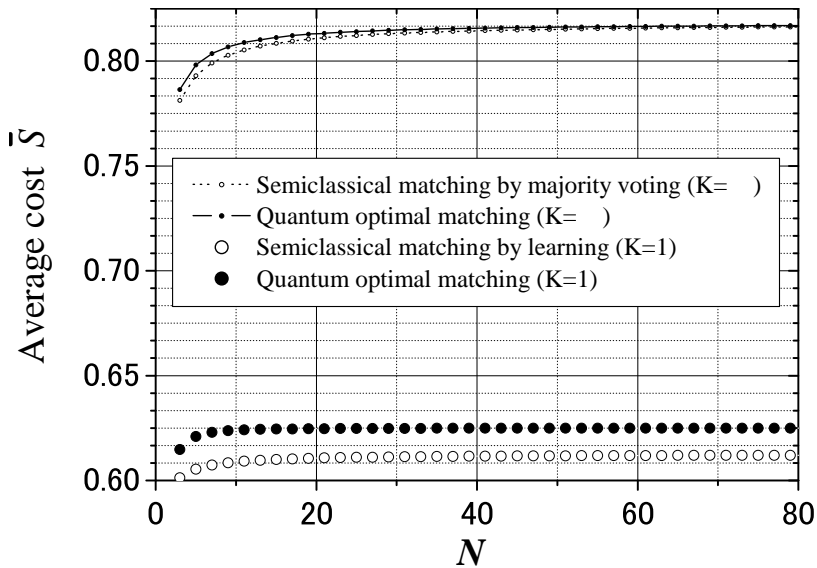}
\end{center}
\caption{\label{fig:ScoreN}
The average score as a function of the number of input feature
systems. The big dots and circles represent the attainable scores by 
the universal quantum matching machine and the optimal 
semiclassical matching strategy based on the learning process, 
respectively, in the case $K=1$. 
The solid and dashed lines are the scores in the case of $K=\infty$ 
derived in our previous paper. 
}
\end{figure}

This score $\bar S^{\mathrm{QM}}$ obtained
by the universal quantum matching machine should be
compared with the optimal score $\bar S^{\mathrm{SC}}$ of the
semiclassical matching strategy based on the learning process. 
Fig. \ref{fig:ScoreN} shows the average score by the two kind of 
matching strategies as a function of $N$, the number of input 
feature samples. 
The big dots represent the average score by the universal quantum 
matching machine in the case of $K=1$. 
This is larger than the one by the optimal 
semiclassical matching strategy based on the learning process, 
shown by the big circle. 
As $K$ increases, we expect larger score although the values can not 
be plotted because we have not succeeded yet in deriving the optimal 
solution for general $K$. 
For $K=\infty$, we have derived the maximum attainable score in 
\cite{Sasaki_Carlini_Jozsa2001a}, 
which is shown by the solid line in Fig. \ref{fig:ScoreN}. 
The dashed line is for the semiclassical matching by majority 
voting. 
%%%%%%%%%%%%%%%%%%%%
%%%%%%%%%%%%%%%%%%%%

\section{Concluding remarks}\label{Concluding}

We have considered a full quantum extension of
the binary quantum pattern matching problem which was addressed
in the recent paper \cite{Sasaki_Carlini_Jozsa2001a}.
In such a problem, given unknown template states
$\ket{g_1}^{\otimes K}$ and $\ket{g_2}^{\otimes K}$, and
an input feature state $\ket{f}^{\otimes N}$,
we are to decide to which template the input feature state is closest
in the sense of the fidelity criterion.
We have presented two kinds of matching strategies, that is,
a semiclassical matching strategy based on learning and
a universal quantum matching strategy.
In particular,
we have explicitly derived the Bayes optimal learning strategy
for the semiclassical matching and the optimal universal quantum
matching strategy
in the case of one copy, $K=1$,
for the template states, and an arbitrary
number $N$ of copies of the input state.
Our previous results in \cite{Sasaki_Carlini_Jozsa2001a}
correspond to the case of $K=\infty$.

For general $K\ge2$,
the Bayes optimal solutions for both the semiclassical
learning strategy and
the universal quantum matching strategy are still not known.
Concerning the optimal learning strategy used in the semiclassical
matching problem,
one of the interesting questions would be whether there
exists an optimal separable strategy of the type as described in
Eq. (\ref{opt_learning_POM_sep}).
  From a preliminary analysis of the case of $K=2$, the POM similar to
Eq. (\ref{opt_learning_POM_sep}), which is now the product
made of two 3-output von Neumann measurements, does not
satisfy the Bayes optimal condition for state estimation.
What would then the optimal learning strategy look like in this case?
Of course there should be a group covariant POM which is generally
an entangled measurement on the two templates.
Is such an entangled measurement
the only optimal learning strategy?
If so, it would be surprising because the two templates have no
a priori correlation.
Or are there other kinds of separable measurements?
As for the universal quantum matching machine,
the problem would just
reduce to finding the appropriate division of the Hilbert space,
but
for larger $K$ this becomes a tedious task.

The reader might feel that
the model used in this paper is in some respect artificial.
In fact,
this model is still far away from practically encountered
situations.
However, we may say that
an important aspect of quantum pattern matching problem is
already seen.
Namely, there certainly exists a full quantum matching procedure as
the universal quantum matching machine which is strictly superior to
the straightforward extension of the conventional matching strategy
based on the learning process of the classifier with the training
template samples.
The derived universal quantum matching machine,
i.e., the POM in Eq. (\ref{optimal POM}), provides a typical
matching model for extracting the meaningful information about
the best template class
without attempting to obtain any further information
about the identities of the template states themselves,
excluding any intermediate measurement process.
In the similar spirit, it is worth mentioning the recent work
on the comparison of two unknown quantum pure states
\cite{BarnettCheflesJex2002}, where the quantum optimal comparing
strategies
are derived for several criteria.

In practical applications, the input and the template systems
will be more complicated, and possibly associated with secondary
features which are not relevant to the pattern classification.
So, as already pointed out in \cite{Sasaki_Carlini_Jozsa2001a},
it would be of practical concern how to enhance the features of
interest and to quarry the essential components (subspaces) of the
quantum system for the pattern classification.
In the scenario where the input and template identities are
completely unknown,
we might rely on a two stage procedure:
first estimate the input and template identities to extract
important features by using some set of aymptotically
vanishing measure of the given samples;
then discard redundant parts of the input and the template
systems, and cut an effective subspace out of the original quantum
Hilbert space;
finally, after the feature enhancement process,
carry out a fully quantum pattern classification procedure in the
smaller space.
Thus, in a sense, we see that the quantum pattern matching problem
naturally involves aspects of
both state estimation and state discrimination.
The former is necessary for the learning process and the feature
enhancement, while the
latter is used for the pattern classification.
In this direction, it would be also interesting to study effective
quantum matching algorithms
which are simple enough in structure and easy to be implemented,
although not necessarily Bayes optimal.

Similarly to the case of the conventional pattern matching problem,
the quantum matching algorithm complexity will be an important
future problem.
It is in fact believed that the complexity in some image recognition
problems is in the NP-complete class.
How can the quantum pattern matching problem be treated from the
point of view of quantum computational complexity?
%The Bayesian approach we adopted here might not be the best ansatz
%for studying the computational complexity of efficient quantum algorithms.
If there will be some progress in the synthesis of a quantum network
for the obtained optimal POM in the Bayesian approach,
then it will be possible to search near optimal quantum matching
algorithms whose complexity might be eventually lower than that
of the corresponding conventional semiclassical approaches.

\acknowledgements

The authors would like to thank R. Jozsa
for helpful discussions. 
%A.C.'s research is supported by ERATO
%under grant no. ????.

\end{document}